# Reservoir Prediction by Machine Learning Methods on The Well Data and Seismic Attributes for Complex Coastal Conditions.

Dmitry Ivlev

Petroleum Overseas ME JSC

dm.ivlev@gmail.com

**The aim of this work** was to predict the probability of the spread of rock formations with hydrocarbon-collecting properties in the studied coastal area using a stack of machine learning algorithms and data augmentation and modification methods. This research develops the direction of machine learning where training is conducted on well data and spatial attributes. Methods for overcoming the limitations of this direction are shown, two methods - augmentation and modification of the well data sample: Spindle and Revers-Calibration. The implementation and effectiveness of these methods are demonstrated on real data.

**Materials and methods.** Data from 15 drilled wells, 930 seismic field attributes and 22 additional space attributes were used in this work. The proposed approach is based on binary classification algorithms with training on well data and includes the following sequence of applying methods: creating training datasets, selecting the features, creating a population of classification models, assessing the quality of the forecast, Reverse-Calibrating the seismic field position, creating additional datasets using Spindle method, assessing the contribution of the features in the forecast, combining models into an ensemble, the final forecast.

**Results.** A prediction of the spatial development of reservoirs was made - a three-dimensional cube of calibrated probabilities of belonging of the studied space to the class of reservoirs was obtained. The estimation of the classification quality for the classification algorithms and changes of the forecast quality from application of the Reversal-Calibration and Spindle methods were done.

**Conclusion.** Considering the difficulties in interpreting seismic data in coastal conditions, the proposed approach is a tool that is capable of working with the entire set of geophysical and geological data, extracting knowledge from a 159-dimensional space of spatial attributes, and making a forecast of the spread of facies with acceptable quality - the F1 measure for the collector class is 0.798 on average for the evaluation of drilling results under different geological conditions. It has been shown that the sequential application of the proposed augmentation methods in the implemented technology stack increases the quality of the collector forecast by more than 1.5 times relative to the original sample.

**Conflict of interest:** Due to a possible conflict of interest, the work lacks geographic referencing, the name of the wells has been changed, and the map does not show the geographic location of the north pole direction.

**Introduction.**

The task of forecasting the spatial distribution of rock formations that are reservoirs for hydrocarbons is both relevant and non-trivial. Machine learning algorithms are one of the tools for such forecasting, and there are different approaches to predicting the geophysical properties of rock formations using them.

This approach has the advantage of training and adjusting models using actual data on environmental properties obtained from geophysical surveys. The complexity of implementing this approach is determined by two constraints: a small sample of well data and uncertainty in the position of the seismic wave field relative to well trajectories. To overcome these constraints, the Spindle method of data augmentation was proposed in [1]: a method of expanding the sample by copying the full sequence of the target geophysical survey or its interpretation with a small shift in the values of spatial data at a new position. This method is based on the assumption that within a given shift in the near-well space, the lithological section does not significantly change, and the sample will include multiple interpretations of the same geological interval.

However, in coastal areas with significant spatial variability of the geological section, large deviations of the well trajectory from the vertical and reduced quality of seismic observations over the area, there are difficulties in predicting changes in environmental properties by machine learning algorithms, including using the proposed data augmentation method.

Based on empirical studies of spatial data in these conditions, the development of the Spindle method - the method of Reverse-Calibrating (RC) the position of the seismic wave field relative to well trajectories - is proposed in this work. This is achieved by searching for the best response of the machine learning model with the target function of reconstructing a continuous sequence of lithology classes along the wellbore using surrounding attribute values in space.

**The aim of this work** is to predict the probability of the presence of rock formations with collector properties in the studied area by implementing a technological stack of machine learning algorithms and demonstrating the implementation of Reverse-Calibration and Spindle methods on actual data, and evaluate the change in forecast quality. The research includes the following sequence of actions: creation of datasets for training, feature selection, creation of a population of classification models, evaluation of forecast quality, creation of an additional dataset using the Spindle method, evaluation of the contribution of features to the forecast, combining models into an ensemble, final forecast, and analysis of the obtained results.

**The study area** is located in the coastal part of a marine basin. A hydrocarbon deposit has been discovered and is being exploited within the study area. The main area of research is located on an exploration license block. Due to the potential conflict of interests, the work does not include a geographical reference to the location, the names of the wells have been changed, and the geographical position of the direction towards the North Pole is not indicated on the map. The initial data consists of a single 3D seismic survey of the area. The measurement overlap ratio decreases from 27 to 9 units towards the coastline. 16 variants of wave field cubes at depths have been obtained as a result of reprocessing using various transformation graphs. An area of study has been extracted from the 16 cubes. The width of the study area is 4875 meters, the length is 11025 meters, and the thickness is 800 meters. The cubes have been brought to a common seismic grid with a lateral step of 25 meters, a vertical step of 4 meters. The total volume of the cube is 17.199 million points. The cubes have a size of approximately 1.1 GB each.

**Primary filtering** of seismic data was carried out based on the Phik correlation coefficient between the values of amplitudes between the cubes. If the coefficient exceeded the threshold of 0.95, one of the cubes was excluded from further work. 5 cubes were left with a Phik coefficient of 0.64-0.92 (Table 1). For each cube, 185 different attributes of the seismic wave field were generated. The space features are standardized and a power transformation Yeo-Johnson is performed.

Table 1. Seismic cubes included in the study

| # | Migration | Route | Stack | Filter | Name in study |
|---|---|---|---|---|---|
| 1 | KPSDM | FWI | BWXT | no | KFRBSU |
| 2 | KPSDM | FWI | FULL | yes | KFRFSU |
| 3 | KPSDM | FWI | IDB | yes | KFRISF |
| 4 | RTM | FWI | BWXT | no | RFRBSU |
| 5 | KPSDM | Tomo | BWXT IDB | yes | KTRBISF |

**An additional set** of features was used in the work, including results from aeromagnetic surveying and lidar imaging, as well as results from stratigraphic interpretation in six versions. The results of the aeromagnetic survey and lidar imaging were projected onto a depth grid with a single value. Discrete cubes with assigned categories between horizons were created based on tracing options for stratigraphic horizons. The total number of additional data was 22 cubes with seismic grid parameters.

**The study used** 15 wells with the results of lithological log interpretation in the study interval: 11 wells from the field and 4 wells from the exploration area. The results of lithological interpretation are coded into two classes - reservoir and non-reservoir. The reservoir is coded as 1, non-reservoir as 0.

**The base dataset** includes results of classification of well logging data projected onto a grid based on the dominant frequency of the class in the seismic cube element intersecting the well trajectory. Additionally, vectors were created from spatial data, where each class definition along the well trajectory was assigned 952 spatial features.

**Feature selection.** Feature selection for further training was carried out in two stages on the base dataset. In the first stage, the BoostARoota algorithm [2] was used. The CatBoost [3] algorithm from the gradient boosting decision tree (GBDT) family was used as an evaluator for its work. In the second stage, the importance of features was evaluated by training the same algorithm, but with the addition of noise to the dataset and calculating the Shapley index for the features. Random data (noise) forms a threshold value for the Shapley index. All features with values of the index below the threshold are excluded from the dataset. After two stages of selection, 159 features of the studied space were left from the original 952 for training. All features from the additional dataset showed low significance for classification and were excluded.

**Augmentation** - expansion and modification of the dataset. Reverse-Calibration was implemented by copying well trajectories along with classification well logging. The copying of wells was done with a large discrete step of lateral displacement of 0, 10, 20 meters to the sides of the world, and up/down in the vertical direction - 0, 5, 10 meters. The total number of trajectories for one well was 125, including the original position. For each trajectory, the lithological well logging classification was projected onto the grid, and vectors were created from spatial data in the position of the shifted trajectory.

**The Reverse-Calibration method** was developed based on two assumptions:

- **first,** the position of the seismic wave field relative to the well trajectory may be shifted due to a more complex physical model of the environment of the studied space compared to the synthetic one;
- **second,** the seismic wave field carries information about the real environment, which implies the existence of an area in the field space whose attributes best correspond to the lithology sequence along the wellbore.

During the exploratory data analysis, the CatBoost algorithm on the test sample showed relatively low-quality classification of collectors on the base dataset (F1 0.64) and even lower quality on the dataset created by the wreath method (F1 0.61). At the same time, experiments with samples showed that when using only part of the data from the wreath method, the quality of the forecast improved (F1 0.66), surpassing the quality obtained on the base dataset.

The search for the position of the seismic wave field relative to the original is carried out by iterating over the positions of the wells in the dataset created using the method of Spindle and optimizing the classification error function of the machine learning algorithm on cross-validation for each combination of trajectories. The positions can be shifted by simultaneously optimizing the error for all datasets, or by optimizing the error for each well independently. The option of simultaneous optimization of all groups of well positions produces a unique combination of well positions

$$X_{d,l} = V(X_{D,L}), d \in D, l \in L,$$

where $X_{D,L}$ is an array of all well positions $D$ generated using the Spindle method, with spatial data vectors $L$, the function $V(\cdot)$ generates $X_{d,l}$ - an array of unique combinations of well $d$ with an array of vectors $l$ for that combination, either randomly or according to a search grid.

$$X_{d,l}^{min} = argmin_{X_{d,l}} [E(C(V(X_{D,L})^n))], n \in N,$$

where each combination $n$ from the number of well position combinations $N$ is given to the classification function $C(\cdot)$, the result of the classification is evaluated by the error function $E(\cdot)$. $X_{d,l}^{min}$ is an array of well sets and their vectors with the minimum classification error value.

This paper implements a version that optimizes the error by shifting all trajectories simultaneously. Given the significant number of combinations of discrete well positions, grid or random search may take a long time, so a search optimizer, the tree-structured Parzen estimator (TPESampler), was used to select combinations [4, 5]. The Random Forest algorithm was used to evaluate the quality of classification.

The quality of classification was evaluated using the ROC AUC metric on cross-validation. Cross-validation was performed on wells, in which data from one well were extracted from the dataset, and the classifier was trained on the remaining data. The quality metric of the trained classifier was assessed on the extracted well, and then the data for the extracted well were returned to the general sample, and the next well was selected from it, and the assessment process was repeated. To evaluate the overall combination of offset wells, the ROC AUC quality assessment results for the wells were averaged. The ROC AUC estimate for the initial positions was 0.627, and for the best-found position 0.810. The distances of the trajectory shifts (shifts seismic wave field) with the best response of the model from the initial are given in Table 2. A trend of the overall trend of the field shift in the area of wells related to the field, and a divergent trend in the search area are noted.

Table 2. Offsets of the seismic field relative to the well trajectories

| Area | Wells | X | Y | Z |
|---|---|---|---|---|
| Field | P-4ST | 10 | -20 | -5 |
| | P-2ST | 10 | -20 | -5 |
| | P-10ST | 10 | -20 | -5 |
| | P-6 | 20 | -20 | 0 |
| | P-3 | 10 | -10 | -5 |
| | P-9 | 10 | -20 | -5 |
| | P-7 | 10 | -10 | -5 |
| | P-8 | 20 | -20 | -5 |
| | P-5 | 0 | -20 | 0 |
| | P-10 | 10 | -10 | -10 |
| | P-11 | 10 | -10 | -5 |
| | **Sub mean** | **10.9** | **-16.4** | **-4.5** |
| | **Sub std** | **5.4** | **5** | **2.7** |
| Wild | J-1X | 20 | 10 | 5 |
| | SEP-1X | -10 | -10 | 10 |
| | SP-1 | 10 | -10 | -5 |
| | SP-1ST | 10 | -20 | -5 |
| | **Sub mean** | **7.5** | **-7.5** | **1.3** |
| | **Sub std** | **12.6** | **12.6** | **7.5** |
| **Total mean** | | **10** | **-14** | **-3** |
| **Total std** | | **7.6** | **8.3** | **4.9** |

Two datasets were formed for further studies: the base dataset (Base) and the Reverse-Calibrated dataset (RC). The number of labels and the ratio of classes are given in Table 3.

Table 3. Quantity of datasets for training

| Class | Base | RC | Spindle RC |
|-------|------|------|------------|
| 0 | 934 | 937 | 57091 |
| 1 | 193 | 192 | 13816 |
| Sum | 1127 | 1129 | 70907 |

**Protocol for data handling** was developed to prevent data leakage - accidental exchange of any information between the test and training sets. The protocol included standardization and Yeo-Johnson power transformation applied to each attribute cube, followed by the transformation parameters being applied to the well datasets (Base and RC) from the same combinations of wells. Eight variations of splits (batches) were formed: training, validation, and test parts for the datasets (Base and RC) from the same combinations of wells. The test parts were formed from data from a combination of two wells: one from the search block and one from the field. These parts were isolated and used only for quality classification assessment. For fine-tuning of models and control of learning, validation parts were formed in the same way - one well from the search part and one from the field. The combination of the eight variations of splits ensured that the data from the test wells would not be included in the training set in any individual variant. At the same time, the eight models of a single machine learning algorithm, in total, saw the entire dataset in various combinations of training and were tuned to the entire range of geological conditions revealed by the wells.

**The design of training** on well data consists of two stages. In the first stage, basic models are trained based on the data processing protocol, after which the models are evaluated for quality on two sets of data for 8 batches. In the second stage, a meta-model is trained on an ensemble of basic models.

**Basic models.** The type of machine learning algorithms currently determines the type of dataset. Data types are divided into unstructured (image, sound, etc.) and structured (tabular). Deep neural network architectures are used for unstructured data. It is believed that the main advantage of deep learning in working with unstructured data lies in the ability to study the feature extraction pipeline (Raschka 2022).

In this work, sets of tabular data were used, where the features of the space - attributes of the seismic field were already extracted and class labels were assigned to them. For such data, gradient boosting decision trees (GBDT) in its specific implementations CatBoost [3], LightGBM [6] and XGboost [7] have proven to be effective in practice.

At the same time, deep learning algorithms for tabular data types are emerging. One such algorithm that showed high classification quality on the datasets used in the work was TabPFN (Prior-data Fitted Network) [8], described as "a trained Transformer that can perform classification". This method is well suited for subsequent ensemble with the results of GBDT algorithms, as its errors are not correlated with the errors of these methods. The current limitation for this algorithm is a maximum of 100 features and a dataset limited to 1000 instances. Therefore, the 100 best features determined using the Shepley index on GBDT family algorithms were input to it. It worked only with non-extended datasets.

**Hyperparameter tuning of models.** For all algorithms, hyperparameter optimization was carried out by maximizing the ROC AUC metric on the cross-validation of the training and validation sets for each batch separately, with balancing of the class ratio and subsequent return of the validation instances to the training set. For gradient boosting decision tree models (GBTD), the metric was maximized using the TPESampler optimizer. For TabPFN, by sequentially iterating from 2 to 200 of the "N_ensemble_configurations" parameter, and the maximum ROC AUC value was reached with this parameter set to 12.

**Model training.** The training of GBDT algorithms was monitored on the validation set for each of the 8 combinations of dataset splits. The TabPFN algorithm, due to its unique characteristics, was trained simultaneously on the validation and training sets. As a result of training four algorithms on the Base and Reverse-Calibrated datasets, 64 models were obtained.

**Evaluating classification quality.** In accordance with the data handling protocol, the quality of classification was assessed on the isolated test set for each batch individually. The F1 metric was used to evaluate the quality of the classes. The metrics were grouped according to the groups being studied and their values were reduced to the mean.

**Classification quality for algorithms.** Table 4 presents the results of evaluating the performance of the algorithms. The average values of the F1 metric for 16 models of each algorithm are shown. The best forecast quality for collectors was demonstrated by the CatBoost algorithm from the GBDT family. The second in quality was the transformer architecture-based algorithm TabPFN. Comparing GBDT algorithms for forecast quality is often a demonstration of the author's expertise in selecting numerous hyperparameters for their settings. This is what distinguishes the TabPFN algorithm, which is trained in one pass on the entire available dataset and currently has one hyperparameter, and at the same time has comparable quality to the GBDT family's forecast. For further research, based on the experimental design, all algorithms used have good forecast quality.

Table 4. Average classification quality (F1 measure) by algorithms

| Parameters | CatBoost | XGBost | LightGBM | TabPFN |
|---|---|---|---|---|
| class 0 | 0.892 | 0.887 | 0.883 | 0.880 |
| class 1 | 0.766 | 0.727 | 0.723 | 0.736 |

**Quality prediction on datasets.** Table 5 shows the averaged prediction metrics grouped by datasets: a dataset with vectors from the original positions of the seismic wave field (Base) and a dataset shifted using the Reverse-Calibration method (RC). It follows that, as a result of using the shifted dataset, the quality of the collector prediction based on the F1 metric increased on average from 0.691 to 0.786. This increase is significant and means that the quality of the collector prediction increased 1.49 times, taking into account the value for a random prediction equal to 0.5. Given that the field calibration by finding the best response to lithology was carried out within the standard error limits of the processing and interpretation results of 3D seismic exploration work, it can be assumed that the new field position better reflects the real lithological sequences exposed by the wells. Thus, it is shown that the method proposed in the work allows to eliminate one of the difficulties of lithological prediction based on seismic data. This can be achieved by separating the field parameters that affect the lithology from the field parameters that affect the physical state of the rock and the properties of the fluid in it.

Table 5. Average classification quality (F1 measure) by datasets

| Parameters | Base | RC | Spindle RC |
|---|---|---|---|
| class 0 | 0.875 | 0.898 | 0.906 |
| class 1 | 0.691 | 0.786 | 0.798 |

**New dataset.** From further research, models and datasets obtained from the original wave field position relative to the well were excluded. From the new positions obtained by Reverse-Calibration, an additional set (Table 3, Spindle RC) was formed using the Spindle method. The discrete copy steps were smaller than for Reverse-Calibration and amounted to 0, 5, 10 meters to the sides of the earth and 0, 2, 4 meters vertically. The number and ratio of the obtained label vectors are given in Table 3. For the new set, according to the data work protocol, a population of models was created and the quality of classification was assessed (Table 5, RC and Spindle RC). The Spindle augmentation method from the new position increased the quality of classification of both classes. The quality increase was 1.56 times for the collector forecast, relative to the original set.

**Feature Importance Assessment.** Using the GBDT model population, the importance of features for forecasting was evaluated (Table 6). The evaluation was carried out for each batch separately using the SHAP library [9] by calculating the Shapley index. The obtained results were scaled (min-max) within the batch, and then the sum was taken for each feature. Table 6 shows 20 out of 159 features with the highest total contribution. The first column of the table shows the name of the seismic cube based on the processing graph used in Table 1, the second column shows the attribute name, and the third column shows the rank by total contribution. The greatest contribution to forecasting was made by attributes of the spectral decomposition with the Morlet wavelet, where the number in the name is the decomposition frequency in Hz. The second most significant for forecasting was the acoustic impedance. The most frequently significant features were obtained from seismic cubes using the Image Domain Beam (IDB) in the processing graph.

Table 6. The twenty most important features for GBDT algorithms

| Seis.cub | Features | Rank |
|---|---|---|
| KFRISF | spectral_dec.-morlet-1 | 1.000 |
| KFRFSU | spectral_dec.-morlet-1 | 0.882 |
| KTRBISF | acoustic_impedance | 0.750 |
| KFRISF | acoustic_impedance | 0.667 |
| KFRISF | spectral_dec.-morlet-3 | 0.590 |
| KFRISF | spectral_dec.-ricker-3 | 0.562 |
| KFRISF | spectral_dec.-morlet-4 | 0.495 |
| KTRBISF | dominant_frequency | 0.477 |
| KFRFSU | amplitude_spectrum | 0.475 |
| KTRBISF | sweetness | 0.462 |
| KTRBISF | spectral_dec.-morlet-2 | 0.402 |
| KTRBISF | spectral_dec.-morlet-1 | 0.391 |
| KFRFSU | spectral_dec.-morlet-4 | 0.361 |
| RFRBSU | sweetness | 0.349 |
| KTRBISF | amplitude_spectrum | 0.321 |
| KTRBISF | spectral_dec.-morlet-13 | 0.316 |
| KFRISF | instantaneous_frequency | 0.292 |
| KTRBSU | spectral_dec.-morlet-1 | 0.288 |
| KTRBISF | instantaneous_frequency | 0.286 |
| KFRFSU | spectral_dec.-ricker-4 | 0.276 |

**Ensemble learning.** For the final prediction of the probability of collectors spread, the obtained model population on the datasets (RC and Spindle RC) was ensembled using the stacking method. This ensemble learning method uses a meta-model that is trained and makes a prediction based on the predictions of the base models. The logistic regression algorithm was used as the meta-model. Before training the ensemble, the GBDT models from the population were further trained on the full dataset - a fixed number of epochs. And for the TabPFN algorithm, an additional model was trained on the full dataset of Reverse-Calibration. The total number of models used for ensemble learning was 57.

**Result of ensemble learning.** As a result of ensemble learning, a prediction was made and a 3D cube of calibrated probabilities of the studied space belonging to the class of collectors was obtained with a seismic data grid resolution (probabilistic space). The space was modified by the volume-interpolated values of the displacement tensor from Table 2 with the opposite sign.

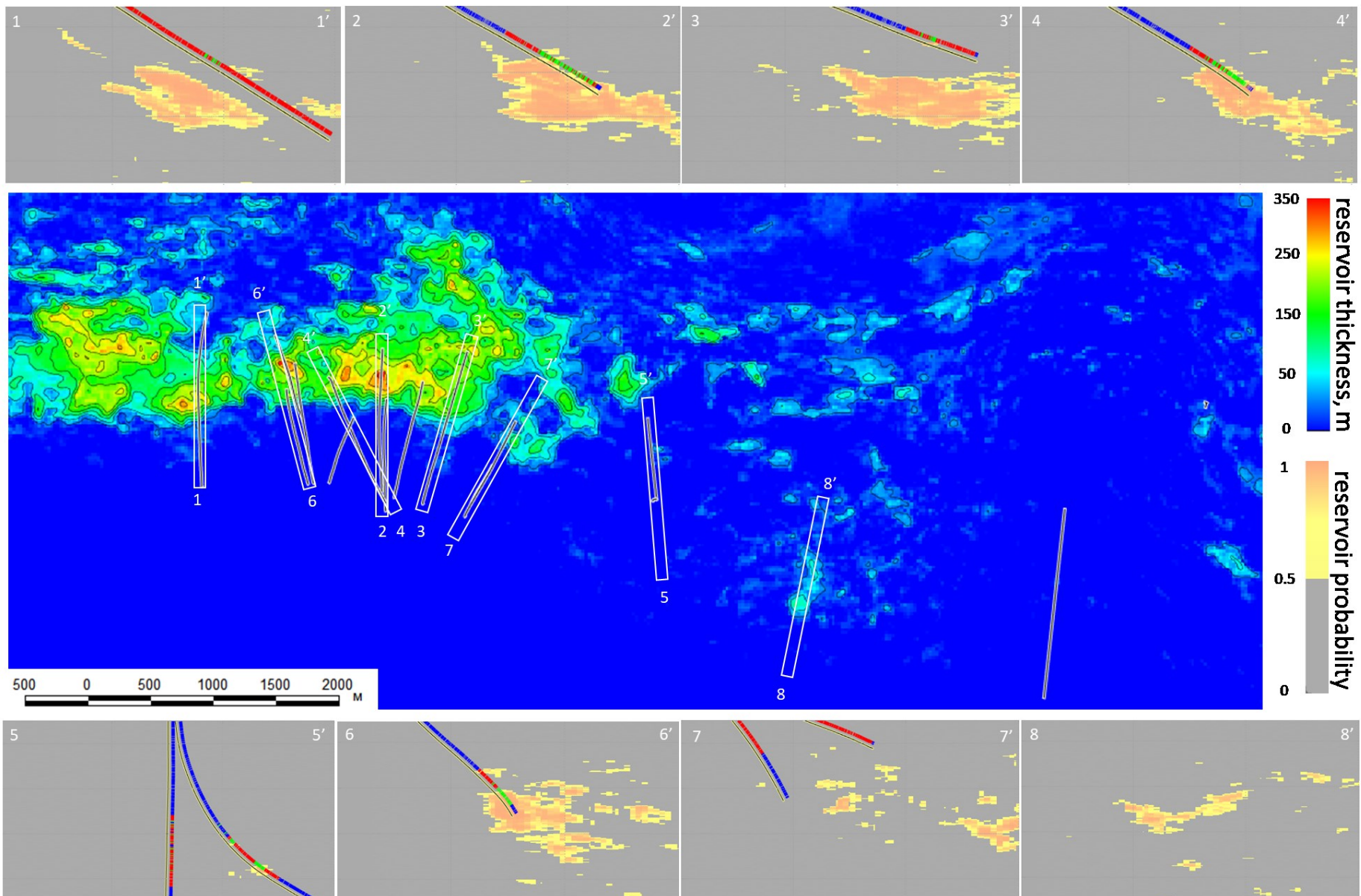

Figure 1. Map of predicted rock thicknesses with reservoir properties within the study area, with white rectangles in the direction of the section. Number – beginning of the section, dashed number – end. The brown lines show the vertiacal projections of the wells. Above and below from the map, vertical sections of the probability space are shown, corresponding to the section numbers from the map

**Description of map and sections.** Figure 1 shows a map of the thickness of rock formations with collector properties within the study area. The map is obtained by vertically summing the voxels of the cube that have a probability of belonging to the collector class greater than 0.5, and multiplying the resulting value by the vertical resolution (4 meters). The sections of the probabilistic space along 7 wells and one prospective area are shown. The sections have a length of 1500 meters and a thickness of 800 meters. Along the well sections, their trajectory and lithological logs are shown with three discrete categories highlighted in colors along the trajectory: green - reservoir, red – non-reservoir, blue - no data.

**Description of the obtained results.** The sections of the probability space 2, 3, 4, 5 show that the model has learned to predict the probability of collectors for well data sufficiently well. The map shows the complex spatial structure of the geological body associated with collectors. The body's presence in the section in the form of steps is well correlated with the geological conception of the development of the research area.

**The final model** is a geology exploration model with a pragmatic task - to find rocks capable of containing hydrocarbons. It was trained on all available data, and its predictive power can only be tested according to the Popper criteria. As an example, to evaluate the model's quality, it is possible to confirm one of the following forecasts using the listed methods:

- drilling exploration well can confirm that there are large traps on the left side of the field with a thickness and area comparable to the open field;
- drilling sidetrack in the lower direction can confirm the presence of part of one of these traps shown on the section in Figure 1;
- drilling sidetrack in the lower direction from the well shown on the third section of Figure 1 can uncover a massive collector;
- drilling exploration well from the shore to an area intersecting section 5 on Figure 1 and not encountering a massive reservoir in the studied interval;
- drilling exploration well from the shore through section 8 on Figure 1 in the area with the maximum predicted probability and encountering deposits related to collectors.

**Conclusions**

1. The complexity of traditional approaches to seismic data interpretation in coastal marine areas is demonstrated on cross-sections 1, 3, 7 in Figure 1. The situation of drilling "blind" wells is characteristic of these conditions, where the historical success rate of even exploitation drilling at the field is 0.5. The proposed approach for these conditions is a tool that, using a technological stack of machine learning algorithms, dataset expansion and modification mechanisms, can work with the entire set of geophysical data, extract knowledge from a 159-dimensional space of seismic attributes, and make a forecast of facies distribution with acceptable quality.
2. It is shown that the average forecast quality for facies belonging to hydrocarbon collectors was F1 0.798 for newly "drilled" wells, always in the test sample, including one well from the field and one well from the search area. The final ensemble of algorithms improved prediction quality.
3. The work develops the direction of machine learning, where algorithms are trained on well data and attribute space. In order to overcome the limitations of this direction, the article develops two methods of well data augmentation:
   - the Spindle method, which enriches the set of well data by information in the near-well space and can be used individually or to implement the Reverse-Calibration method;
   - the Reverse-Calibration method, which searches for a new position of the seismic wave field relative to well trajectories by finding the best response of the machine learning model with the target function of restoring a continuous sequence of lithology classes along the wellbore using surrounding attribute space values.

During the implementation of the entire technological stack of machine learning algorithms, it was shown that the sequential application of the augmentation methods - Spindle and Reverse-Calibration, proposed in the work, increases the forecast quality of collectors for this research area by 1.56 times relative to the original dataset.

## List of references